\documentclass[11pt,twoside]{article}
\usepackage{asp2010}

\resetcounters

\begin{document}

\title{Momentum-driven feedback and the \boldmath $M$-$\sigma$
  relation in non-isothermal galaxies} 

\author{Rachael C. McQuillin and Dean E. McLaughlin
  \affil{Astrophysics Group, Lennard Jones Laboratories, Keele
    University, Keele, Staffordshire, ST5 5BG, UK}} 

\begin{abstract}
We solve for the velocity fields of momentum-conserving supershells
driven by feedback from supermassive black holes or nuclear star
clusters (central massive objects: CMOs).  We treat, for the first
time, the case of CMOs embedded in gaseous protogalaxies with
non-isothermal dark matter haloes having peaked circular-speed
profiles.  We find the CMO mass that is 
sufficient to drive {\it any} shell to escape {\it any} such halo.
In the limit of large halo mass, relevant to real galaxies,
this critical CMO mass depends only on the peak circular speed in the
halo, scaling as $M_{\rm crit} \propto V_{\rm c,pk}^4$.
{\null\vskip-0.5pt}

\end{abstract}

\section{Introduction}

The $M$--$\sigma$ relation between supermassive black hole (SMBH) mass
\citep[e.g.,][]{gultekin_09} or nuclear star cluster (NC) mass
\citep{ferrarese_06} and the stellar velocity dispersion of galaxy
bulges can be understood as a result of momentum-conserving feedback
\citep[e.g.,][]{king_05,mcl_06}.  Recently, \citet{volonteri_2011} have
argued that SMBH mass also correlates with the asymptotic
circular speed at large radii in bulges. Such an $M$--$V_{\rm{c}}$
relation should presumably also be related to feedback, a point that
we investigate in \citet{mcq_2012} and summarize here.

Accretion at near- or super-Eddington rates onto an SMBH in a gaseous
protogalaxy is expected to result in a fast wind driven back into the
galaxy. Similarly, the 
winds and supernovae from massive stars in a young NC will
drive a superwind into its host galaxy.  In a spherical approximation
to either case, the wind from the central massive object (CMO) sweeps
ambient gas into a shell that, at least initially, cools rapidly
and is therefore momentum-driven \citep{king_03,mcl_06}.
For a CMO larger than some critical mass, the wind thrust
($\propto\!M_{_{\rm{CMO}}}$) can overcome the gravity of the host dark
matter halo (measured by $\sigma$) and the shell can escape, cutting off
fuel to the CMO and locking in an $M_{_{\rm CMO}}$--$\sigma$ relation
and associated scalings.  

\citet{king_05} showed that the CMO mass required to drive a
shell at large radius in a singular isothermal sphere with velocity
dispersion $\sigma_0$ is 
\begin{equation}
M_{\sigma}  ~\equiv~  f_0  \kappa \sigma_0^4 \,\big/\,
(\pi  G^2 \lambda) ~=~
4.56 \times 10^8 ~ M_{\odot} ~ \sigma_{200}^4 \, f_{0.2} \,
\lambda^{-1} ~~, 
\label{eq:msig}
\end{equation}
where $f_0$ is a gas mass fraction (assumed to be spatially constant,
and with $f_0 \approx 0.2$ so $f_{0.2}=f_0/0.2$); $\kappa$ is the
electron scattering opacity; and $\sigma_{200} = \sigma_0 / 200
\,\mathrm{km\,s^{-1}}$.  The parameter 
$\lambda$ is related to the feedback efficiency of each CMO type:
$\lambda \approx 1$ for SMBHs, and $\lambda \approx 0.05$ for NCs
\citep{mcl_06}. In \citet{mcq_2012}, we derive a version of
eq.~(\ref{eq:msig}) for
{\it non-isothermal} dark matter haloes. In this more realistic and
general case, the critical CMO mass depends on a
{\it peak circular speed}. 

\section{Velocity Fields of Momentum-Driven Shells}
We assume that the CMO wind thrust is $dp_{\rm{wind}}/dt=\lambda
L_{\rm{Edd}}/c$ \citep{king_pounds_03,mcl_06}.  The equation of motion
for a momentum-driven shell in any dark matter halo profile
$M_{_{\rm{DM}}}(r)$ is therefore \citep[see also][]{king_05} 
\begin{equation}
\frac{d}{dt} \left[ M_{\rm{g}}(r)v \right] ~=~  
         \frac{4 \pi G \lambda M_{_{\rm{CMO}}}}{\kappa} 
           - \frac{GM_{\rm{g}}(r)}{r^2}
           \left[M_{_{\rm{CMO}}}+M_{_{\rm{DM}}}(r)\right] ~~. 
\label{eq:eqmotion1}
\end{equation}
Here $r$ is the instantaneous radius of the shell; $v=dr/dt$ is the
velocity of the shell; and $M_{\rm{g}}(r)$ is the ambient gas mass
originally inside radius $r$ (i.e., the mass that has been swept up
into the shell when it has radius $r$).  Then, letting $M_{\rm{g}}(r)
\equiv f_0 \, h(r) \, M_{_{\rm{DM}}}(r)$, where $h(r)$ is a function
describing how the gas traces the dark matter [i.e., $h(r)\equiv 1$
  when the gas traces the dark matter directly], we re-write
eq.~(\ref{eq:eqmotion1}) to solve for the velocity field $v(r)$ rather 
than $r(t)$ directly: 
\begin{equation}
\frac{d}{dr} \left[M_{_{\rm{DM}}}^2 h^2 v^2(r) \right] ~=~ 
  \frac{8 \pi G \lambda}{f_0 \kappa} M_{_{\rm{CMO}}}h(r)M_{_{\rm{DM}}}(r) -
    \frac{2M_{_{\rm{DM}}}^2(r)h^2(r)}{r^2} \left[M_{_{\rm{CMO}}} +
      M_{_{\rm{DM}}}(r) \right]  ~~.
\label{eq:eqmotion}
\end{equation}

\section{Singular Isothermal Sphere}

\begin{figure}
\plotone{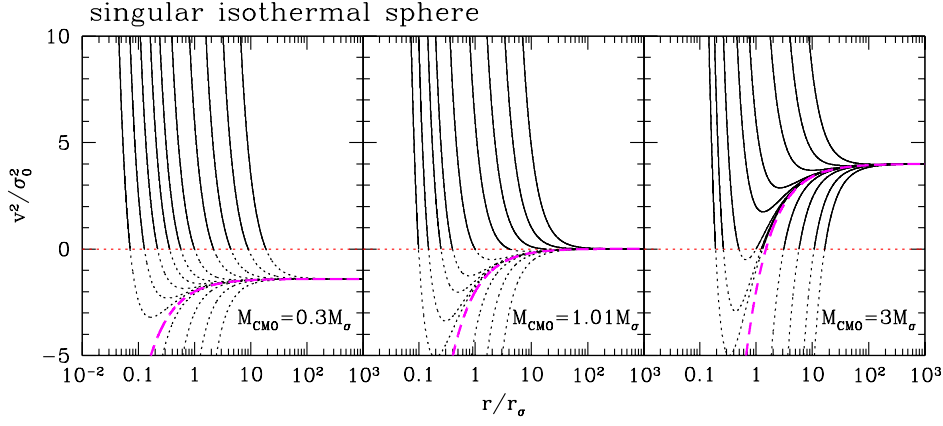}
\caption{Velocity fields, $v^2$ versus $r$, for momentum-driven shells
  in a singular isothermal sphere with spatially constant gas
  fraction (eq.~[\ref{eq:sis_sol}]).
  The long-dashed curve in each panel is the solution with $C=0$ for
  the CMO mass indicated. The physical ($v^2\ge 0$) parts of solutions
  with $C \neq 0$ are shown as solid lines. The mass unit $M_\sigma$ is
  defined in eq.~(\ref{eq:msig}); the radius unit is
  $r_{\sigma} \equiv GM_{\sigma}/\sigma_0^2 \simeq 49 ~\mathrm{pc}~
  \sigma_{200}^2$.} 
\label{fig:sisvsq}
\end{figure}

We first revisit the singular isothermal sphere, in which
$M_{_{\rm{DM}}}(r)=2\sigma_0^2 r/G$. With this mass profile and $h(r)
\equiv 1$, so the gas traces the dark matter directly,
eq.~(\ref{eq:eqmotion}) has solution  
\begin{equation}
v^2 ~=~ 2\sigma_0^2 \left[ \frac{M_{_{\rm{CMO}}}}{M_{\sigma}} -1 \right]
       - \frac{2GM_{_{\rm{CMO}}}}{r}
          + \frac{C}{r^2} ~~,
\label{eq:sis_sol}
\end{equation}
where $M_{\sigma}$ is defined in eq.~(\ref{eq:msig}) and the
constant of integration, $C$, represents the shell momentum at
$r=0$.  This solution is only physical if $v^2>0$.  At arbitrarily
large radius, the shell tends to a constant coasting speed,
$v_{\infty}^2 \equiv 2\sigma_0^2 \left[ M_{_{\rm{CMO}}}/M_{\sigma} -1
  \right]$, so that $M_{_{\rm{CMO}}}>M_{\sigma}$ is required for
shells to move at all at large $r$. This is fundamentally the result of
\citet{king_05}.  But at 
small radius, the term $C/r^2$ in eq.~(\ref{eq:sis_sol}) dominates and
the behaviour of a shell is determined by its initial momentum. 

Figure \ref{fig:sisvsq} plots $v^2$ versus $r$ from
eq.~(\ref{eq:sis_sol}) for $M_{_{\rm{CMO}}}/M_{\sigma}=$0.3, 1.01 and
3, with a range of $C$ values in each case.   
When $M_{_{\rm{CMO}}}<M_{\sigma}$ (left panel), no shell can ever
escape.  All solutions either have the shell stalling (i.e.,
crossing $v^2=0$) at a finite radius, or are unphysical ($v^2<0$) at
all radii.  
When $M_{_{\rm{CMO}}}=1.01M_{\sigma}$ (middle panel), only a few
solutions formally allow shells to reach large $r$ without
stalling. However, these potential escapes require initial ($r=0$)
shell momenta, $C$, so large as to give 
velocities $v \ga 0.2\,c\,\sigma_{200}$ at radii
$r\sim 1~{\rm pc}~\sigma_{200}^2$; and even then, they tend to a
large-$r$ coasting speed of just $v_{\infty}=0.14\,\sigma_0$.
Thus, $M_{_{\rm{CMO}}}>M_{\sigma}$ is a
{\it necessary but not sufficient} condition for the escape of
momentum-driven feedback from an isothermal sphere.
For $M_{_{\rm{CMO}}}=3M_{\sigma}$ (right panel), the ability to reach
large radii still
depends on the initial momentum of the shell.   
Shells decelerating from small
radius require $v \ga 0.05\,c\,\sigma_{200}$ at
$r \sim 1~{\rm pc}~\sigma_{200}^2$ to avoid
stalling. Launch solutions (those accelerating outwards from $v^2=0$
at a non-zero radius) are also possible, but only
starting from radii $r\ga 40~{\rm pc}~\sigma_{200}^2$.

In the case $M_{_{\rm CMO}}=3M_\sigma$, the shells that
reach large radii eventually coast at $v_{\infty} = 2 \sigma_0$,
which is the escape speed from an isothermal sphere. We
conclude that at least $M_{_{\rm{CMO}}} \geq 3M_{\sigma}$ is
required (along with a large shell momentum at $r=0$) for
escape. The objection by  
\citet{silk_nusser_2010} to momentum-conserving black hole winds as
the sole source of the SMBH $M$--$\sigma$ relation ultimately traces
back to our result for $v_\infty^2$.
However, {\it the problem is specific to the assumption of an
isothermal dark matter halo}.

\section{Non-Isothermal Dark Matter Haloes}

More realistically, dark matter haloes have density profiles that are
shallower than isothermal at small radii and steeper than isothermal
at large radii.
The circular speed, $V_{\rm{c}}^2=GM_{_{\rm DM}}(r)/r$,
corresponding to any such density profile increases outwards from the
centre, has a well-defined peak, and then declines towards larger
radii. As a result, momentum-driven shells always accelerate at large
radii \citep{mcq_2012} and are guaranteed to exceed the halo
escape speed eventually, just so long as they do not stall before
reaching radii where they can begin to accelerate.

\begin{figure}
\plotone{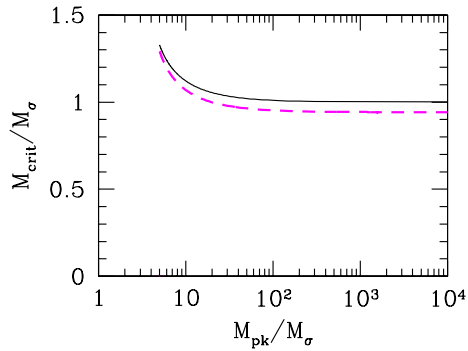}
\caption{{\it Solid line}: the CMO mass (in units of $M_{\sigma}$
  defined by eq.~[\ref{eq:mcrit}]) that is {\it sufficient} for the
  escape of {\it any} momentum-driven shell from an NFW halo with dark
  matter mass $M_{\rm pk}$ inside the peak of its circular speed
  curve. {\it Dashed line}: the {\it necessary} CMO mass for the
  escape of shells with initial momentum zero {\it specifically}.}
\label{fig:nfwmex}
\end{figure}

The natural velocity unit in such non-isothermal haloes is the peak
value of the circular speed, $V_{\rm c,pk}$, and we define the
associated mass
\begin{equation}
M_{\sigma} ~\equiv~ \frac{f_0 \kappa}{\lambda \pi G^2}\,
\frac{V_{\rm{c,pk}}^4}{4} 
~=~ 1.14 \times 10^8 ~M_{\odot}~
\left(\frac{V_{\rm{c,pk}}}{200\, \mathrm{km\,s^{-1}}} \right)^4
f_{0.2} \, \lambda^{-1} ~~.
\label{eq:mcrit}
\end{equation}
An obvious choice of fiducial velocity dispersion here is
$\sigma_0\equiv V_{\rm c,pk}/\sqrt{2}$, for which
eq.~(\ref{eq:mcrit}) reduces to eq.~(\ref{eq:msig}).
In either form, given our results for the isothermal sphere above and
for other halo models below (e.g., Figure \ref{fig:nfwmex}),
$M_\sigma$ is best viewed as just a unit. 

A general analysis of equation (\ref{eq:eqmotion}) shows that, in
any dark matter halo with a single-peaked circular-speed curve,
there is a critical CMO mass, $M_{\rm crit}$, that {\it is sufficient}
to guarantee the escape of shells with {\it any initial momentum}
\citep{mcq_2012}. This $M_{\rm crit}$ depends primarily on
$V_{\rm c,pk}$ and on the dark matter mass, $M_{\rm pk}$, within the
peak of the circular-speed curve. In the limit that
$M_{\rm pk}\gg M_\sigma$, which is most relevant for real galaxies,
we find that $M_{\rm crit}$ tends to a value that is
independent of any other details (i.e., the exact shape) of the dark
matter mass profile:
\begin{equation}
M_{\rm{crit}} ~\simeq~
M_{\sigma} \left[1 \,+\, M_{\sigma}\big/M_{\rm{pk}}
                  \,+\, \mathcal{O}
                  \left(M_{\sigma}^2\big/M_{\rm{pk}}^2\right)\right]
~\longrightarrow~ M_\sigma ~~.
\qquad\quad
(M_{\rm pk} \gg M_\sigma)
\label{eq:res}
\end{equation}

\begin{figure}
\plotone{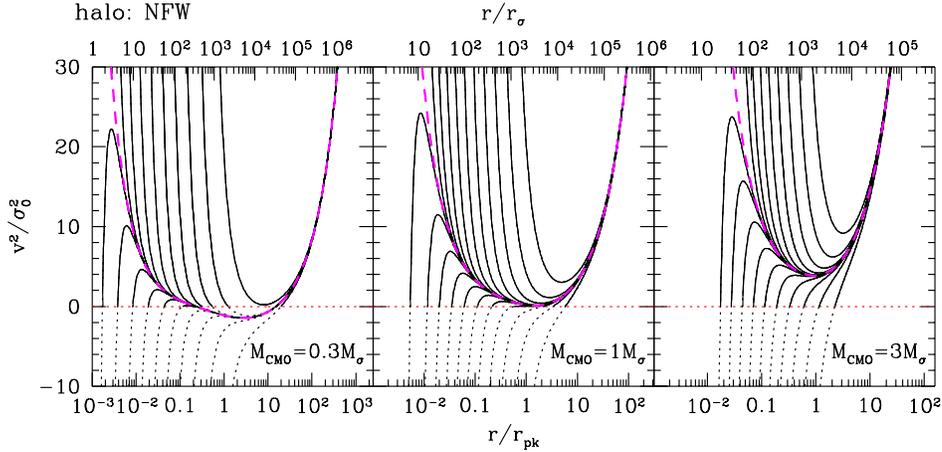}
\caption{Velocity fields for momentum-driven shells in a Milky
  Way-sized, NFW halo with spatially constant gas fraction.
  Long-dashed curves represent solutions with zero momentum at $r=0$.
  Velocities are put in units of
  $\sigma_0\equiv V_{\rm c,pk}/\sqrt{2}$. The mass unit $M_\sigma$ is
  defined in eq.~(\ref{eq:mcrit}). The radius unit on the lower axis
  is the radius at which the dark matter circular-speed curve peaks;
  in the Milky Way, $r_{\rm{pk}} \approx 50~{\rm kpc}$. On the upper
  axis, $r_{\sigma}\equiv GM_\sigma/(V_{\rm c,pk}^2/2)
        \simeq 24~{\rm pc}~(V_{\rm c,pk}/200~{\rm km~s}^{-1})^2$.}
\label{fig:nfwvsq}
\end{figure}

We have applied our general analysis to the specific dark matter halo
models of \citet{hernquist_1990},
\citet[][NFW]{nfw_97}, and \citet{dehnen_mcl_05}.
Figure \ref{fig:nfwvsq}, from \citet{mcq_2012},
shows numerical
solutions of eq.~(\ref{eq:eqmotion}) for the velocity fields of
momentum-driven shells in an NFW halo with parameters
appropriate to an $L_\star$ galaxy: a dark matter circular-speed
curve peaking at  $r_{\rm pk} = 50$~kpc with
$V_{\rm c,pk}= 200~{\rm km~s}^{-1}$, and thus a dark matter mass of
$M_{\rm pk} \simeq 4.7\times10^{11}~M_\odot$
inside $r_{\rm pk}$. Equations (\ref{eq:mcrit}) and
(\ref{eq:res}) then give
$M_{\rm crit}\simeq 1.00024\,M_\sigma$ for the critical CMO mass.

The left panel of Figure \ref{fig:nfwvsq} shows $v^2$ versus $r$ for
shells with different initial momenta, given
$M_{_{\rm CMO}}=0.3\,M_\sigma$.
Shells that decelerate from large velocity at
small radius hit $v^2=0$ and stall, unless they have an impossible
$v\ga 10^6\,c$ at $r\sim1~{\rm pc}$.
Shells launched from radii $r<r_{\rm pk}$ go on to
stall at some larger radius, which is still inside $r_{\rm pk}$.
Launches from $r>r_{\rm pk}$ accelerate monotonically outwards
and therefore 
always escape, but they only ever start from implausibly large
$r \ga 500\,$kpc.  
By contrast, when $M_{_{\rm{CMO}}}=M_{\sigma}$ (middle panel), all the
solutions shown are able to escape, regardless of their
initial velocities or momenta---as expected, since
$M_{\rm crit}$ is so near $M_\sigma$ here.
The right-hand panel of the figure confirms
that all shells escape for larger CMO masses.

The solid line in Figure \ref{fig:nfwmex} shows the CMO mass
$M_{\rm crit}$  that is sufficient for the escape of {\it any}
momentum-driven shell from an NFW halo, as a function of the halo mass
$M_{\rm{pk}}$. As expected from eq.~(\ref{eq:res}),
$M_{\rm crit}\rightarrow M_\sigma$ for large $M_{\rm pk}\gg M_\sigma$.
The dashed line in the figure shows the CMO mass that is
{\it necessary} for the escape of shells with zero momentum at $r=0$
{\it specifically}. This is slightly smaller than the sufficient
CMO $M_{\rm crit}$ at any halo mass, tending to a value of
$\simeq\!0.94\,M_\sigma$ for $M_{\rm pk}\gg M_\sigma$.

\section{Conclusion}

The critical CMO mass of King (2003, 2005; eq.~[\ref{eq:msig}]) is
necessary but {\it not sufficient} for the escape of momentum-driven
feedback from {\it isothermal} protogalaxies. 
By contrast, in {\it non-isothermal} dark matter haloes with realistic,
single-peaked circular-speed curves, equations 
(\ref{eq:mcrit}) and (\ref{eq:res}) above give a CMO mass that
{\it is sufficient} for the escape of {\it any} momentum-driven
shell: 
$M_{\rm crit}\simeq 1.14\times10^8 ~M_\odot~
 (V_{\rm c,pk}/200~{\rm km~s}^{-1})^4 \,f_{0.2}\,\lambda^{-1}$,
if the haloes are massive
(essentially, $M_{\rm pk}\gg M_{\rm crit}$) and 
have spatially constant gas mass fractions.
Our full analysis can be found in \citet{mcq_2012}.

Our results directly predict an $M_{_{\rm CMO}}$--$V_{\rm{c,pk}}$
relation, and thus provide a basis from which to address claimed
relations between SMBH mass and asymptotic circular speed in bulges
\citep[e.g.,][]{volonteri_2011}. 
The characteristic galaxy velocity dispersion that is relevant to
observed $M_{_{\rm CMO}}$--$\sigma$ relations is then
$\sigma_0\equiv V_{\rm c,pk}/\sqrt{2}$---although it is an open issue
how this relates generally to measured velocity dispersions within the
{\it stellar} effective radii of bulges.
Meanwhile, working in terms of $V_{\rm{c,pk}}$ seems a natural way
to extend the analysis of galaxy--CMO correlations to systems
with significant rotational support as well as pressure support.
This could be of particular interest in connection with NCs in
intermediate-mass early-type galaxies, and even in bulgeless Sc/Sd
disks.



\begin{thebibliography}

\bibitem[\protect\citeauthoryear{Dehnen \& McLaughlin}{2005}]{dehnen_mcl_05}
 Dehnen, W., \&  McLaughlin, D. E. 2005, MNRAS, 363, 1057
\bibitem[\protect\citeauthoryear{G\"ultekin et al.}{2009}]{gultekin_09}
 G\"ultekin et al. 2009, ApJ, 698, 198
\bibitem[\protect\citeauthoryear{Ferrarese et al.}{2006}]{ferrarese_06}
 Ferrarese et al. 2006, ApJ, 664, 17
\bibitem[\protect\citeauthoryear{Hernquist}{1990}]{hernquist_1990}
 Hernquist, L. 1990, ApJ, 356, 359
\bibitem[\protect\citeauthoryear{King}{2003}]{king_03}
 King, A. R. 2003, ApJ, 596, L27
\bibitem[\protect\citeauthoryear{King}{2005}]{king_05}
 King, A. R. 2005, ApJ, 635, L121
\bibitem[\protect\citeauthoryear{King \& Pounds}{2003}]{king_pounds_03}
 King A. R., \& Pounds, K. A. 2003, MNRAS, 345, 657
\bibitem[\protect\citeauthoryear{McLaughlin et al.}{2006}]{mcl_06}
 McLaughlin, D. E., King, A. R., \& Nayakshin, S. 2006, ApJ, 650, L37
\bibitem[\protect\citeauthoryear{McQuillin \& McLaughlin}{2012}]{mcq_2012}
 McQuillin, R. C., \& McLaughlin, D. E. 2012, MNRAS, submitted
\bibitem[\protect\citeauthoryear{Navarro et al.}{1997}]{nfw_97}
 Navarro, J. F., Frenk, C. S., \&  White, S. D. M. 1997, ApJ, 490, 493
\bibitem[\protect\citeauthoryear{Silk \& Nusser}{2010}]{silk_nusser_2010}
 Silk, J., \& Nusser, A. 2010, ApJ, 725, 556
\bibitem[\protect\citeauthoryear{Volonteri et al.}{2011}]{volonteri_2011}
Volonteri, M., Natarajan, P., \& G{\"u}ltekin, K. 2011, ApJ, 737, 50
\end{thebibliography}
\end{document}